%% file: main.tex
\documentclass[10pt,twocolumn,letterpaper]{article}

\usepackage[pagenumbers]{cvpr} %
\input{preamble}

\definecolor{cvprblue}{rgb}{0.21,0.49,0.74}

\usepackage[pagebackref,breaklinks,colorlinks,citecolor=cvprblue]{hyperref}

\newcommand{\highlightnewtext}{\color{black}}

\newenvironment{hlbox}
{
\begin{tcolorbox}[enhanced,attach boxed title to top center={yshift=-3mm,yshifttext=-1mm},
  colback=blue!5!white,colframe=blue!75!black,colbacktitle=white,coltitle=black,fonttitle=\bfseries,
  boxed title style={size=small,colframe=blue!75!black}]
}
{
\end{tcolorbox}
}

\makeatletter
\renewcommand{\paragraph}{%
  \@startsection{paragraph}{4}%
  {\z@}{1.25ex \@plus 1ex \@minus .2ex}{-1em}%
  {\normalfont\normalsize\bfseries}%
}
\makeatother

\usepackage[utf8]{inputenc}
\usepackage[T1]{fontenc}

\usepackage{enumitem}
\setlist{nosep}

\def\paperID{18} %
\def\confName{CVPR}
\def\confYear{2024}

\title{Towards a GENEA Leaderboard -- an Extended, Living Benchmark for Evaluating and Advancing Conversational Motion Synthesis}

\author[1]{Rajmund Nagy}
\author[2]{Hendric Voss}
\author[3]{Youngwoo Yoon}
\author[4]{Taras Kucherenko}
\author[5]{Teodor Nikolov}
\author[6]{\\{}Thanh Hoang-Minh}
\author[7]{Rachel McDonnell}
\author[2]{Stefan Kopp}
\author[8]{Michael Neff}
\author[1,5]{Gustav Eje Henter}
\makeatletter
\renewcommand\AB@affilsepx{, \protect\Affilfont}
\makeatother

\affil[1]{KTH Royal Institute of Technology}
\affil[2]{Bielefeld University}
\affil[3]{ETRI}
\affil[4]{SEED -- Electronic Arts}
\affil[5]{Motorica AB}
\affil[6]{University of Science -- VNUHCM}
\affil[7]{Trinity College Dublin}
\affil[8]{University of California, Davis}
\begin{document}

\twocolumn[{
\maketitle
    \centering
    \captionsetup{type=table}
    \begin{tabular}{@{}ll@{}}
    \toprule
    \textbf{Shortcomings of gesture-generation evaluations} & \textbf{How the GENEA Leaderboard may address them} \\
    \midrule
    \multirow[t]{2}{*}{Infrequent comparisons between leading models} & Maintain a living benchmark with recurring human evaluations \\
    & Release a collection of SOTA model outputs \\\rule{0pt}{3ex}\multirow[t]{3}{*}{Low user-study standardisation and reproducibility} & Develop a reproducible evaluation standard \\
    & Provide automated scripts for crowd-sourced evaluations \\
    
    & Provide an easy-to use, open-source 3D-visualiser tool \\
    \rule{0pt}{3ex}\multirow[t]{2}{*}{Lack of a definitive benchmark dataset} & Determine the most suitable dataset for standardised  benchmarking \\
    & Improve dataset quality and annotations \\
    \rule{0pt}{3ex}\multirow[t]{2}{*}{Misaligned evaluation tasks} & 
    Record and promote new datasets with highly grounded gestures \\
    & Adopt task-driven, in-context evaluation for the leaderboard\\%Define new modelling tasks and evaluation settings \\
    \bottomrule
    \end{tabular}
    \captionof{table}{We are developing a community-driven holistic solution addressing the major challenges in evaluating gesture-generation models.}
    \label{tab:evaluation_roadblocks}
\vspace{0.5\baselineskip}
}]

\begin{abstract}
Current evaluation practices in speech-driven gesture generation lack standardisation and focus on aspects that are easy to measure over aspects that actually matter. This leads to a situation where it is impossible to know what is the state of the art, or to know which method works better for which purpose when comparing two publications.

In this position paper, we review and give details on issues with existing gesture-generation evaluation, and present a novel proposal for remedying them. Specifically, we announce an upcoming living leaderboard to benchmark progress in conversational motion synthesis. Unlike earlier gesture-generation challenges, the leaderboard will be updated with large-scale user studies of new gesture-generation systems multiple times per year, and systems on the leaderboard can be submitted to any publication venue that their authors prefer. By evolving the leaderboard evaluation data and tasks over time, the effort can keep driving progress towards the most important end goals identified by the community.

We actively seek community involvement across the entire evaluation pipeline: from data and tasks for the evaluation, via tooling, to the systems evaluated. In other words, our proposal will not only make it easier for researchers to perform good evaluations, but their collective input and contributions will also help drive the future of gesture-generation research.

\end{abstract}

\vspace{-1\baselineskip}
\section{Introduction}    
Our face-, body-, and hand movements serve important communicative functions when we speak, or listen to others. Embodied agents like social robots and human-like game characters therefore should also employ such \emph{co-speech gestures} when communicating. However, manually animating conversations, or even just choosing and timing pre-recorded gesture clips for some given speech, are challenging and time-consuming tasks. Consequently, the last decade has seen a long line of data-driven models developed for automatic gesture synthesis~\cite{chiu2015predicting, yoon2020speech, kucherenko2020gesticulator,li2021audio2gestures,liu2022learning, ahuja2022low, liang2022seeg, pang2023bodyformer, habibie2022motion}, including recent applications of diffusion models~\cite{zhu2023taming,ao2023gesturediffuclip, alexanderson2023listen, yang2023diffusestylegesture,deichler2023diffusion}, which have achieved breakthrough results in other generative AI domains. %

\highlightnewtext
In this position paper, we examine recent deep learning-based gesture-generation models through the lens of evaluation. How much have these models improved over recent years? What are the strongest models today in terms of complementary aspects like motion quality and speech-gesture alignment? Our analysis shows that gesture-generation evaluations are generally inconsistent and not directly comparable to each other, thus it is impossible to answer the above questions. We argue that a coordinated, community-driven effort is needed to unify the fragmented evaluation landscape, and propose a solution (\cref{tab:evaluation_roadblocks}):
\color{black}

\begin{hlbox}
We announce a community-driven project for developing the \textbf{GENEA Leaderboard}\footnotemark, the first living benchmark for gesture generation. 
\end{hlbox}
\footnotetext{We build on the name of the GENEA gesture-generation challenges~\cite{kucherenko2023genea} to highlight the similar goals of the two initiatives. The acronym stands for ``Generation and Evaluation of Non-verbal behaviour for Embodied Agents''.}
Additionally, we contrast the limited range of contemporary evaluation questions with the variety of communicative functions gestures serve, and advocate for recording and utilising new, more contextualised speech-gesture datasets, as they enable high-impact research directions.

The paper is organised as follows:
\begin{enumerate}
    \item A survey and a critical perspective on the evaluation practices of data-driven gesture generation (\cref{sec:current_state}).
    \item The announcement and design of the GENEA Leaderboard (\cref{sec:leaderboard}), aimed to address the ongoing limitations of gesture-generation evaluation.
    \item A collection of potential research directions towards practically applicable gesture-generation models, facilitated by the leaderboard project (\cref{sec:new_directions}).
\end{enumerate}

\begin{hlbox}
We are actively looking for feedback on the leaderboard proposal, its implementation, and its evolution. If you are reading this position paper and you have an idea on how we can improve the GENEA Leaderboard, please email us at \href{mailto:genea-leaderboard@googlegroups.com}{genea-leaderboard@googlegroups.com}!    
\end{hlbox}

\input{reviewed_works}

\section{Limitations of gesture-generation evaluation}
\label{sec:current_state}
\highlightnewtext
A key question in gesture-generation research is: \textit{What is currently the best gesture-generation model?} More specifically, we might ask: \textit{Which model performs best in areas like human-likeness, or matching gestures to speech, or controlling style?} These questions are important, but we can only answer them if the field follows solid evaluation practices.

In this section, we review how recent data-driven gesture-generation studies approach evaluation. As a comprehensive survey is beyond the scope of this paper (for that, see \cite{wolfert2021review, nyatsanga_comprehensive_2023}), we focus our analysis on 3D gesture-generation research presented at selected leading graphics and vision conferences from 2022 onward. Using the search terms ``gesture'', ``co-speech'', ``speech'', and ``motion'' in publication titles, we identified a total of 24 relevant works, as listed in \cref{tab:recent_models}.

Our findings, detailed in the rest of this section, %
show that contemporary results in gesture generation are not directly comparable, and most papers lack the careful set-up necessary for disentangling complementary notions of gesture quality. The GENEA Leaderboard, described in \cref{sec:leaderboard}, is our proposal for a sustained evaluation effort that directly tackles the highlighted challenges.

\subsection{Objective evaluation}
\label{sec:surveyed_object_evaluation}
Automated, ``objective'' metrics are essential in deep learning: in the form of loss functions, they enable us to train models; in possibly other forms, they facilitate scalable evaluation. Unfortunately, defining useful, holistic metrics is generally a challenge for generative machine-learning problems, and this is especially true for gesture generation.

Each surveyed paper (\cref{tab:recent_models}) includes some combination of objective metrics for quantitative performance analysis. Some of them are based on the mean error of the data (e.g., \textbf{PCK} \cite{yang2011articulated}, \textbf{SRGR} \cite{liu2022beat}, or the $\ell_2$-norm of joint position/rotation/acceleration). These metrics treat motion synthesis as a regression problem, and are therefore fundamentally flawed, as gesturing is widely known as an especially idiosyncratic and non-deterministic type of human movement \cite{mcneill2000language}. Other evaluations \cite{sun2023co, liu2022learning, zhi2023livelyspeaker, chhatre2024amuse} feature \textbf{beat consistency} metrics originating from dance synthesis \cite{li2022danceformer, li2021learn}. While these heuristics might reflect speech-gesture alignment to a degree, their failure modes are not well understood and they only describe a small aspect of good gesturing.

By far the most prominent objective metric is the Fréchet Gesture Distance (\textbf{FGD})~\cite{yoon2020speech}, which is an application of the FID score~\cite{heusel2017gans} to motion data. FID scores have been instrumental in the development of neighbouring fields, e.g., image and video generation, but the validity of the metric hinges on the feature extractor model and dataset it is trained on. For example, in image generation, the powerful Inception-V3 image classifier \cite{szegedy2016rethinking}, trained on the large-scale ImageNet dataset \cite{deng2009imagenet}, has served for several years as the de facto encoder for acquiring representations for computing the FID scores. Unfortunately, gesture generation does not have a similarly standardised large-scale motion-capture dataset, or a perceptually relevant discriminant task like image classification. Some gesture-generation papers use the autoencoder provided in \cite{yoon2020speech}, but this model is trained on the Human3.6M dataset \cite{ionescu2013human3}, which contains very limited range of gesturing behaviour. Instead, authors often train their own feature extractors without conducting any assessment of the perceptual relevance of the learnt representations. Furthermore, even if the same feature extractor is used, metrics such as FGD still cannot be compared between papers if aspects such as sample size differ \cite{chong2020effectively} (and in practice, they often do).
\color{black}
Overall, the degree and nature of alignment between the above objective metrics and human perception of gesturing is an understudied topic. However, recent experiments validating objective metrics of gesture quality \cite{yang2023qpgesture, kucherenko2024evaluating} found that none of the included metrics displayed a strong correlation with human opinion. Mismatches between similar objective metrics and human preference have also noted by other motion synthesis works \cite{tseng2023edge, dabral2022mofusion}. We conclude that objective metrics in gesture generation are inconsistently applied, and their validity is not sufficiently established in the literature.
\color{black}
\begin{hlbox}
Using humans for gesture assessment -- subjective evaluation -- must thus be considered the gold-standard evaluation methodology in gesture generation, and is the key focus of this paper.
\end{hlbox}

\subsection{Baselines and datasets}\label{ssec:surveyed_datasets_baselines}

\begin{figure}
    \highlightnewtext
    \centering
    \includegraphics[width=0.8\linewidth]{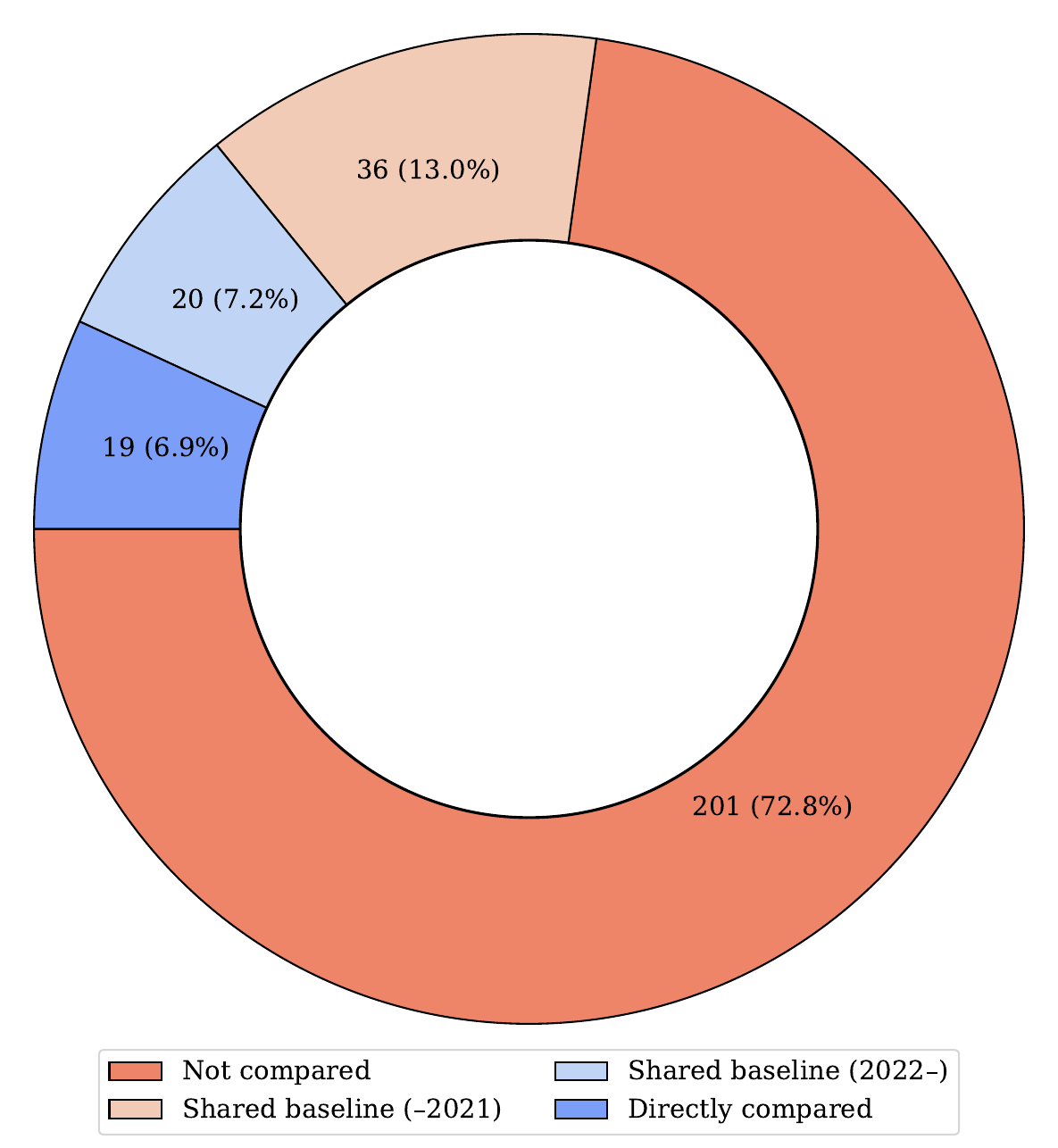}
    \caption{A breakdown of the availability of human evaluation results for all 276 possible pairings of the 24 models listed in\cref{tab:recent_models}. For a given model pair $(a,b)$, ``Directly compared'' means that the authors of $a$ compare to $b$ or vice versa; ``Shared baseline'' means that there is no direct comparison, but there is a third model from \cref{tab:recent_models} that both the authors of $a$ and $b$ compare to (with the publication date of the latest such baseline in parentheses); ``Not compared" means that there is no direct comparison, nor a shared baseline, using the previous definitions.}
    \label{fig:comp-breakdown}
\end{figure}
\highlightnewtext
Almost all papers in \cref{tab:recent_models} include subjective evaluation among their results. In fact, several papers stress the importance of subjective evaluation due to the perceived unreliability of objective metrics (e.g., \cite{pang2023bodyformer, ao2023gesturediffuclip, zhang2024SemanticGesture, yang2023qpgesture, ng2024audio2photoreal}).

However, evaluations can only effectively measure progress if they include strong, up-to-date baseline models. Ideally, the network of comparisons (via human evaluations) made between models over recent years would form a graph with a few well-connected clusters, each representing a sub-problem that has been thoroughly evaluated. 

To evaluate whether this is the case for gesture generation, we ask the following question for all 276 possible pairings of models from \cref{tab:recent_models}: ``\emph{What relevant human evaluation results are available in the publications of either model?}'', with the results visualised on \cref{fig:comp-breakdown}. 

For the majority of cases, the unfortunate answer is ``\emph{None}''. Evaluations against shared baselines are available for 55 out of 276 possible model pairs. In 36 cases, the shared baselines were published 2021 or earlier -- these can likely be considered ``weak'' baselines, therefore comparisons against them are insufficient to differentiate between two potentially state-of-the-art models. In 20 cases, a shared baseline dated 2022 or later is available, which can serve as a more informative point of reference. However, in order to assess the relative differences between each model and the baseline, the individual user studies need to have comparable setups. As we will show in \cref{sec:surveyed_subjective_evaluation}, this is generally not the case. 

Direct comparisons were only performed for 19 pairings (just below 7\% of the 276 cases), as visualised in the last column of \cref{tab:recent_models}. These evaluations are the strongest form of evidence for ranking two models, and they will be the focus of our GENEA Leaderboard, as described in \cref{sec:leaderboard}.

The main reason for the relatively low occurrence of direct comparisons between recent models is the great variation in datasets used. The most frequently used datasets for the 24 models are BEAT \cite{liu2022beat} (8 cases), TED \cite{yoon2019robots} (6 cases), the Trinity Speech-Gesture Dataset \cite{ferstl2018investigating,ferstl2021expressgesture} (5 cases), in addition to 16 other datasets used in 3 publications or fewer. Due to this variability, there is a limited set of available pre-trained baselines for a given user study. Unfortunately, without reliable objective metrics, the best results are often achieved by selecting model weights and tuning the sampling parameters (e.g., temperature or decoding scheme) based on visual inspection of the generated motion. Therefore, properly adapting a baseline from one dataset to another is often unfeasible. As a result, reimplemented or newly trained baselines often fail to match the qualitative properties of the original models, e.g., they show jittery motion \cite{pang2023bodyformer, chen2024diffsheg}. This detracts from the reliability of evaluations.%
\color{black}
\begin{hlbox}
    It is currently impossible to determine the state of the art in gesture generation. Recent models are seldom compared to each other or to a common, informative baseline using human evaluation. This is unlikely to change without concerted effort due to the many datasets in use and the difficulty of adapting previous models to new data. 
\end{hlbox}

\subsection{Subjective evaluation}
\label{sec:surveyed_subjective_evaluation}
Although subjective evaluation can take many shapes, by far the most common gesture-motion aspects to be evaluated are the \emph{human-likeness} of the motion and the gesture \emph{appropriateness for the speech}.

\paragraph{Human-likeness} The first type of evaluation is intended to quantify to what extent gestures visually look like something a human might produce. A typical approach is to show video stimuli of gesture motion and ask participants to rate their perceived human-likeness on a numerical or Likert-type scale. Words like ``quality'', ``naturalness'', ``realism'', and ``lifelikeness'' might be used interchangeably in literature and in the task description to the raters. Sometimes, this evaluation is paired with a separate user study on motion smoothness in order to identify jerky, still, or lethargic motion, which are common failure modes of poorly trained gesture models.

\paragraph{Appropriateness for speech} This second type of evaluation is intended to quantify the link between the gestures and the speech, i.e., how well gestures are grounded in the content of the speech. Whilst information used to ground gestures can theoretically take many forms (see \cref{sec:new_directions} and \cite{nyatsanga_comprehensive_2023}), current evaluations only tend to consider the semantic content of gesturing, its alignment to rhythm and prosody, or both.  

\subsubsection{Implementation factors} \label{ssec:implementation_factors}
\highlightnewtext
In \cref{ssec:surveyed_datasets_baselines}, we have identified 20 pairs of models where the authors do not make a direct comparison between the pair, but do compare to a shared baseline that is also included in \cref{tab:recent_models}. However, it turns out that these indirect comparisons can seldom be used to ascertain the ranking between the pairs of models due to differences in user-study design and execution. To show this, we take a closer look at the 12 publications (marked with a blue square in \cref{tab:recent_models}) that compare to another model included in our analysis.

The \textbf{types of collected human ratings} include forced pairwise preference votes, forced orderings, preference votes with ties allowed, 5-point Likert-style preference votes, and 5-point mean opinion score (MOS). Similarly, the \textbf{question formulations} are not standardised; a few examples are listed in \cref{tab:example_question_formulations}. Whilst the \textbf{duration of the generated motion clips} is mostly standardised between 8-15 seconds, we find that authors do not specify the exact clips they select, and \textbf{stimulus videos are never released} across the surveyed papers. Moreover, the \textbf{demographics of recruited evaluators} is rarely reported, and the best practice of including attention checks is not consistently applied.

Unfortunately, there is evidence from neighbouring fields, like text-to-speech evaluation, that these kinds of subtle design choices can have a significant effect on human ratings and system rankings \cite{chiang2023why, kirkland2023stuck}, as can the choice of baseline systems \cite{cooper2023investingating}. Numerical scores are thus not comparable across papers.

We also observe differences in the \textbf{embodiment} used: 4 studies use 2D skeletons, 3 studies use the SMPL-X mesh without additional texture, 3 studies use distinct 3D meshes, and one study does not include details about the embodiment. The realism of the visualised character as well as its compatibility with the speech and the motion can have a major influence on the sensitivity of the ratings (as shown by, e.g., \cite{ng2024audio}).

\begin{table*}[]
    \centering
    \begin{tabular}{ll}
        \toprule
         User-study category & Example questions and instructions \\
         \midrule
         \multirow{4}{*}{Human-likeness} & Which motion is most believable? \cite{liu2024emage} \\
         & Evaluate the realism and naturalness of the holistic expression and gestures. \cite{chen2024diffsheg} \\
         & Which of the two gesture motions appear more natural? \cite{mughal2024convofusion} \\
         & Does the motion resemble a real human? \cite{ao2023gesturediffuclip} \\
         \midrule
         \multirow{4}{*}{Appropriateness for speech} & Assess the rhythmic coherence between gestures and speech audio. \cite{zhang2024SemanticGesture} \\
         & Which of the two gesture motions corresponds better with spoken utterance? \cite{mughal2024convofusion}\\
         & Which video has better synchronisation of gestures and speech rhythm? \cite{chhatre2024amuse}\\
         & How appropriate are the gestures for the speech? \cite{yang2023qpgesture} \\
         \bottomrule
    \end{tabular}
    \caption{Example user-study question formulations, highlighting the lack of standardisation between user studies.}
    \label{tab:example_question_formulations}
\end{table*}

\color{black} 
\begin{tcolorbox}[enhanced,attach boxed title to top center={yshift=-3mm,yshifttext=-1mm},
  colback=blue!5!white,colframe=blue!75!black,colbacktitle=white,coltitle=black,fonttitle=\bfseries,
  boxed title style={size=small,colframe=blue!75!black}]
We conclude that subjective evaluation methods in gesture generation generally have low reproducibility, and their results are impossible to directly compare to each other.
\end{tcolorbox}

\subsection{Related work: the GENEA Challenges}
\label{sec:genea}
The GENEA Challenges \cite{kucherenko2021large, yoon2022genea, kucherenko2023genea} are the sole effort in providing a systematic, fair comparison for a large selection of gesture-generation models. The challenges are structured as time-limited events, organised at three conferences so far (at ACM IVA in 2020, and at ACM ICMI in 2022 and 2023). The organisers provide a cleaned-up mocap dataset with pre-determined training-test splits, and a rendering script for creating videos from model outputs in a pre-determined neutral scene. 
Participating teams have around 6--8 weeks to develop their models and submit their synthesised output for the entire test set. From these larger set of synthetic motion clips, organisers manually select a number of short clips for the evaluation, which prevents submitting cherry-picked results. To create the user-study stimuli, the selected segments of the dataset and the submitted synthetic data are rendered into videos, shown either one at a time, or with two videos in parallel, depending on the study. Finally, the organisers conduct a large-scale crowdsourced human evaluation of all submitted systems, one or two baselines, and the reference mocap data, and perform statistical analyses of the collected human ratings.

\begin{table}[!t]
    \centering
    \begin{tabular}{@{}ccccc@{}}
    \toprule
    Year & Dataset & Models& Eval.\ clips & Test-takers \\
    \midrule
        2020&Trinity \cite{ferstl2018investigating}&\hphantom{0}7&40&\hphantom{0}250  \\
        2022&TWH \cite{lee2019talking}&11&48&\hphantom{0}822 \\
        2023&TWH \cite{lee2019talking}&14&70&1223 \\
    \bottomrule
    \end{tabular}
    \caption{Key statistics for the GENEA challenge evaluations.}
    \label{tab:genea_stats}
    \vspace{-1\baselineskip}
\end{table}

The primary goal of the GENEA Challenges is to directly compare a large number of gesture-generation models under controlled conditions, with an order-of-magnitude larger number of test-takers than what has been typical in gesture generation (see \cref{tab:genea_stats}). However, an equally important contribution of the challenges lies in their continuously evolving evaluation methodology, described below.

\subsubsection{Disentangling human-likeness and appropriateness to speech}
\label{sec:disentangling}
Although it is standard practice to
evaluate the complementary aspects of \emph{human-likeness} and \emph{appropriateness for speech} of synthetic gestures (see \cref{sec:surveyed_subjective_evaluation}), there is little reason to believe that typical evaluations successfully disentangle the two. In contrast, the GENEA Challenges take careful measures to separate the two evaluations.

In particular, the human-likeness evaluations are intended to be purely visual, and speech content is therefore omitted from stimuli (so no audio nor subtitles) in all three challenges, in order to evaluate only the motion. This decision is partly based on earlier results in facial motion synthesis, which showed that the presence of a speech track can improperly influence human ratings \cite{jonell2020letsfaceit}. 

The first GENEA Challenge \cite{kucherenko2021large} assessed human-likeness and appropriateness using an identical methodology (HEMVIP \cite{jonell2021hemvip}), and merely opted to change the evaluation question on the user-study interface. This closely mirrors the evaluation practices of recent models listed in \cref{tab:recent_models}. Very interestingly, \emph{mismatched videos} playing the selected speech segments but showing unrelated motion clips from the test set surpassed all synthetic gesture conditions in terms of appropriateness rating \cite{kucherenko2021large}. This strongly indicates that models with high human-likeness scores may receive relatively high appropriateness scores even when they do not depend on the speech. Therefore, it is crucial to control for the overall naturalness of the motion when conducting appropriateness evaluations. The latter GENEA challenges leverage \emph{mismatching} \cite{ennis2010seeing} to propose a new evaluation paradigm to achieve this \cite{yoon2022genea, kucherenko2023genea}. 

The core idea of the mismatching is to present stimulus pairs where one contains motion that matches the speech, whereas the other contains motion from the same source (so similar average human-likeness) but unrelated to the speech presented in the video. Participants are then asked to indicate which video within the pair has motion that best matches the speech. If a model generates motion that is semantically or rhythmically linked to the content of the speech, then it is reasonable to assume that evaluators will tend to prefer the stimulus with matched motion on average. On the other hand, if a model does not depend on the speech or otherwise fails to capture the links between speech and the gesture, the average evaluator preference between the matched and mismatched clips should be close to neutral (although any differences in human-likeness between matched and mismatched motion segments may still have an impact on participant responses \cite{kucherenko2024evaluating}). This more controlled methodology has identified a substantial appropriateness gap between natural motion from 3D mocap and synthetic gesture motion \cite{yoon2022genea,kucherenko2023genea}.

Whilst the GENEA challenges combine several notions of gesture appropriateness under a single evaluation question, the mismatching paradigm can be easily adapted for studying more detailed questions. Unfortunately, mismatching is not often used the works of \cref{tab:recent_models}. As a result, although some papers report appropriateness results that closely approximate or even surpass their measured appropriateness of human mocap \cite{yang2023qpgesture, yi2023generating, zhu2023taming}, they can not be considered evidence for near-human appropriateness, having made no attempt to control for the very substantial confounding effect of motion human-likeness.

\subsubsection{Outcomes of the GENEA Challenges}

Overall, the GENEA challenges are leading contributions to gesture-generation evaluation for several reasons:
\begin{itemize}
    \item Providing the only large-scale subjective evaluations of contemporary gesture models; 
    \item Improving evaluation practices: common visualisation (used in \cite{wang2021integrated,teshima2022deep,zhang2023diffmotion}), better subjective evaluation methodologies (used in \cite{yoon2021sgtoolkit}), and benchmarking against the challenge submissions \cite{ferstl2021expressgesture, yazdian2022gesture2vec}.
    \item Providing strong empirical evidence of the unsuitability of most objective metrics \cite{kucherenko2024evaluating}.
    \item Lowering barriers to entry by providing post-processed datasets, visualisation scripts, subjective evaluation tools \cite{jonell2021hemvip}, and ready-made stimuli to compare to.%
\end{itemize}
However, we argue that, ultimately, \textbf{the GENEA challenges do not achieve their goal of comprehensively benchmarking the state of the art in gesture generation}. Due to the short timeline of the challenge format, evaluations mainly feature models developed under severe time pressure, and the requirement to publish participating systems at the host venue (historically ACM IVA and ACM ICMI) is a limitation that may discourage authors from participating.

\section{Towards a new standardised benchmark}
\label{sec:leaderboard}
\highlightnewtext

\input{roadblocks_sections}
Our analysis of evaluation practices has highlighted several roadblocks for reproducible development and evaluation of gesture-generation models (\cref{tab:roadblocks_sections}). We argue that a new community-driven effort is needed in order to unify the fragmented evaluation landscape of gesture generation with simultaneous goals of:
\begin{enumerate}
    \item Creating a repository of standardised resources and tooling;
    \item Conducting recurring large-scale systematic comparisons of recently published gesture-generation models in collaboration with model authors;
\end{enumerate}
and we describe our proposed GENEA leaderboard, designed to address all of the above points. 

\subsection{Benchmark dataset}\label{sec:benchmark_dataset}
In order to construct the leaderboard, we need to designate at least one benchmark dataset, and select evaluation segments from the test set for the human evaluation. Both of these are of central importance to the initiative, since dataset characteristics affect both the capabilities of participating systems and the behaviour of human evaluators.

\subsubsection{Deciding on the dataset}
Currently, 3D motion-capture technology remains the best way of acquiring high-quality gesturing data \cite{nyatsanga_comprehensive_2023}, especially for hand and finger information. Among currently available public mocap datasets \cite{nyatsanga_comprehensive_2023}, we think the \textbf{BEAT2 dataset} \cite{liu2024emage} is the most promising candidate for standardised benchmarking thanks to its large size and high variety of speakers and emotions. \highlightnewtext Furthermore, this dataset is a refined version of BEAT \cite{liu2022beat}, which is the most commonly used dataset in \cref{tab:recent_models}, and its official baseline CaMN is the most common baseline among the user studies we analysed in \cref{ssec:surveyed_datasets_baselines}. Finally, unlike most other datasets, BEAT2 captures high-quality facial animations in addition to the body motion, which can increase the realism of the user-study visualisations, and will enable us to evaluate the joint generation of face- and body gestures.

Successfully narrowing the focus of the gesture-generation community to fewer datasets might increase the scientific value of (and therefore foster) dataset augmentations like additional text- and gesture annotations; motion-capture cleanup; or additional recordings. 
\subsubsection{Dataset format}
Unlike the datasets used in the GENEA challenges, which were based on the BVH (BioVision Hierarchy) file format, BEAT2 provides data in SMPL-X \cite{pavlakos2019cvpr} format.
Whilst BVH is commonly used in the animation industry and much of the human-computer interaction community, SMPL-X is becoming increasingly widely used by gesture generation researchers, especially in publications at machine-learning and computer-vision conferences.
We believe this is because SMPL-X is easier to use for researchers and engineers, since it provides a canonical human mesh that does not require rigging or other animation expertise to animate.
Unlike BVH, SMPL-X inherently decouples body-shape parameters from pose parameters, which simplifies learning a joint model using data from different persons with different body and skeleton proportions.
(In other words, the user does not have to perform any motion retargeting).
In addition, SMPL-X integrates with the FLAME \cite{li2017learning} representation of facial motion.
As BEAT2 includes captures of facial motion in the FLAME format, this provides a straightforward way to incorporate facial motion into the leaderboard in the future.

Although a proprietary format, SMPL-X and its surrounding toolchain is freely available for research, similar to how the data in recent GENEA challenges only permitted non-commercial use.
We therefore currently believe it is the best choice for starting out the leaderboard initiative.

\highlightnewtext
\subsubsection{Determining evaluation segments}\label{ssec:segment_selection}
As we highlighted in \cref{ssec:surveyed_datasets_baselines}, the standard practice in gesture generation is to use randomly selected short segments for evaluation, which we believe to be far from ideal.

To maximise the reproducibility and the transparency of the leaderboard evaluations, our plans are the following:
\begin{enumerate}
    \item We will create official train-test splits based on recommendations from the authors of BEAT2.
    \item We will divide the test split into roughly one-minute \emph{segments}.
    \item Each submission to the leaderboard will have to include several (around 3\textendash5) synthetic outputs for each segment. (This will allow us to test the stochasticity of the systems.)
    \item For objective evaluation, we will use all submitted outputs.
    \item We will manually select around 50\textendash100 short (7\textendash15 second) \emph{clips} from the segments, which will be published for reproducibility. The timestamps for these clips will be shared publicly, and their selection criteria will ensure that they do not end or begin in the middle of a sentence; that semantically meaningful gestures are highly represented; that there are no unusual artefacts present, and that the speaker is active. 
    \item For human evaluation, we will randomly choose one of the submitted variations for each clip. 
\end{enumerate}

Overall, we believe that these criteria will significantly affect the efficacy of the evaluations, promote methods that can generate long sequences without quality deterioration, and prevent  cherry-picked outputs to a large degree.

\subsection{Recurring human evaluation}
\label{sec:recurring_evaluation}
\begin{figure}[!t]
    \centering\includegraphics[width=0.95\linewidth]{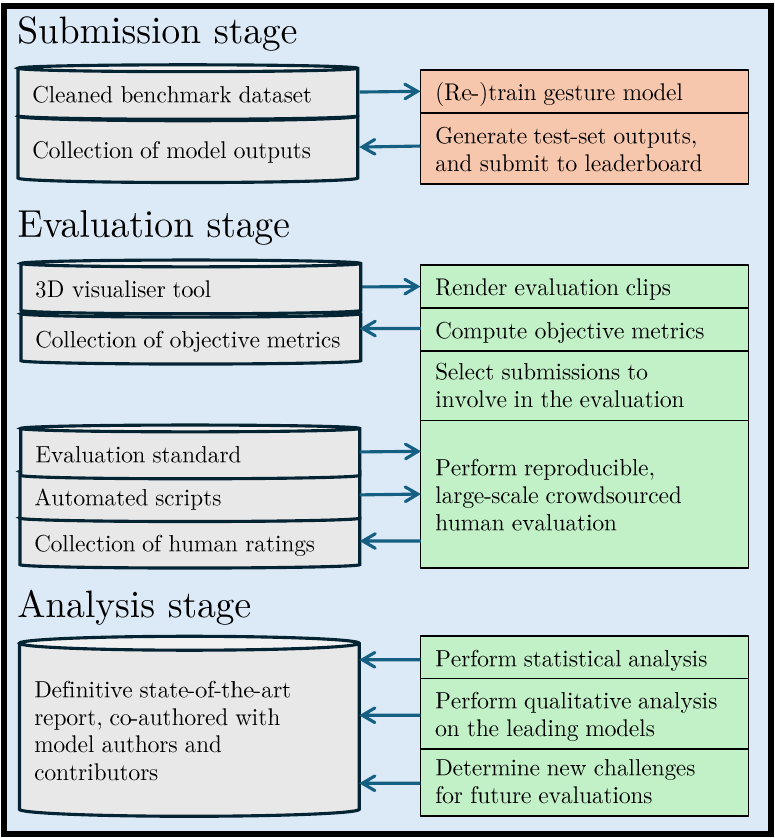}
    \caption{An overview of the resources provided by the leaderboard (left side); the steps involved in the recurring evaluation (right side), with organiser tasks in green and author tasks in peach colour; and how these connect to each other (arrows).}
    \label{fig:leaderboard}
    \vspace{-1\baselineskip}
\end{figure}

The core innovation of the GENEA Leaderboard concept is to recognise the need for a \emph{living benchmark format} that involves recurring human evaluation performed by a group of maintainers (with responsibilities described \cref{fig:leaderboard}). Our proposal constitutes a ``living'' human-evaluation benchmark, both due to the more rapid updates compared to events like the GENEA challenges, and because the evaluation data and tasks will evolve with the continuing updates of the leaderboard. This is important to refine the evaluation approaches and to keep driving progress towards meaningful research targets in gesture generation.

\subsubsection{Overall setup}
In contrast to the fixed annual timelines of the GENEA challenges, the leaderboard will feature a \textbf{rolling submission system}, with evaluations taking place continuously. 
\begin{hlbox}
    We envision that, once a sufficient level of automation has been achieved, the time between motion submission and results being available will be on the order of one or two weeks.
\end{hlbox} 
This includes the time required for manual verification of submissions and possible dialogue and clarifications with submitting teams.
Evaluation results and other data about submitted systems will be made available in a leaderboard format, as is common in ML and computer vision, but has not been done in gesture generation. %

\highlightnewtext
\subsubsection{Initial seeding of models}\label{ssec:initial_seeding}
Our initial aim is to establish a definitive ranking of many of the most promising contemporary gesture generation-models (including but not limited to the works of \cref{tab:recent_models}) through centralised evaluation. %

In order to include definitive versions of leading gesture-generation models into the leaderboard at launch, we will invite the original authors of these models to re-train their respective models on the BEAT2 dataset in the SMPL-X format. This is necessary because model tuning is currently a very manual and undocumented process (partly due to the lack of reliable objective metrics), meaning that a model re-trained by a third party may not be considered reliable. We believe that this is a great opportunity for collaboration between the leaderboard organisers and other model developers, as we will fully fund and conduct the evaluations, and we will invite participating authors to co-author an upcoming state-of-the-art report based on the evaluation results. 

\highlightnewtext
\subsubsection{Participation criteria}
After the initial seeding is done, we will open up the leaderboard to the public for new submissions. Unlike the GENEA challenges, submissions will not be constrained to publish at specific conferences, making it easier to cater to and bring together multiple communities.
However, participants will still be required to provide documentation (e.g., a link to a paper or a technical report and filling in a form with questions about aspects such as training data, model size, and real-time factor) in sufficient detail to understand and build on the model, and put their results in context.
This information can also allow sorting and filtering the results on the leaderboard, e.g., to only compare real-time systems or systems that were trained on specific datasets.

Due to the monetary cost as well as the time involved in conducting evaluations, we will likely have to filter submissions. To this end, we intend to use certain objective metrics (e.g., FGD score, jerkiness) for preliminary evaluation, but ultimately, whether a system is allowed to participate will be determined by manual inspection of the generated outputs, and possibly smaller-scale user studies. Our general principle is that models that would clearly be in the bottom half of the rankings will not be allowed to participate, but exceptions may be made for systems with unique features (e.g., real-time generation, or systems trained on low amounts of data) on a case-by-case basis.

\subsubsection{Ranking system}

Compared to the one-off evaluations in the GENEA Challenges, a rolling submission system imposes new requirements on the evaluation methodology. For example, the number of ratings between systems can greatly differ, and it will be necessary to continuously integrate new results from user studies that might have non-overlapping sets of systems and stimuli.
 
We believe an Elo-like ranking system is a natural choice for the Leaderboard, since it is an established ranking method based on pairwise comparisons that can handle the above challenges. Additionally, thanks to the popularity of the approach, Elo scores are generally well understood and will be a more transparent form of presenting user-study results. 

Following the methodology of Chatbot Arena \cite{zheng2024chatbot_arena} -- another large-scale human evaluation effort -- we are going to use the Bradley-Terry (BT) model \cite{Zermelo1929, bradley1952rank} to establish our rankings. Preliminary experiments on the GENEA 2023 user-study data indicate that BT scores with bootstrapped 95\% confidence intervals have higher separation between top-ranking systems, and yield statistically significant rankings with lower number of evaluations -- both of which are important properties for our Leaderboard. In future work, we are planning to perform detailed experiments comparing different ranking systems, and develop adaptive stimulus- and system-selection methods to further increase the efficiency of the evaluation.

\subsection{User-study design and tooling}\label{sec:user_study_design}
The initial evaluation to seed the leaderboard with results will be based on the GENEA-Challenge methodology, with two evaluation questions (human-likeness and appropriateness for speech).
In order to ensure consistency and reliability of repeated user studies, we will develop automation for running crowdsourced evaluations. Furthermore, an important part of standardisation will be to create detailed instructions on aspects of the study that might influence the results, including the attention check protocols, evaluator recruitment process, the time of day and duration of the study (which may affect response quality), and so on.

As for user-study tooling, we are currently working on an enhanced version of the HEMVIP codebase \cite{jonell2021hemvip}, originally used in the GENEA challenges. Our primary focus is to improve the system’s usability while introducing automation features through API integrations with crowd-sourcing platforms such as Prolific. Making standardised automation tools and procedures publicly available will vastly lower the threshold for other researchers to perform best-practices subjective evaluations of their own systems against the latest methods as baselines, e.g., for pilot studies.

\subsection{Visualisation tooling}\label{ssec:visualisation_tooling}
As gesture animations are intended for human viewers, it is important to inspect input data and model output during both model development and evaluation. 

In motion synthesis, data is typically stored as a sequence of joint rotations, specific to a certain skeleton structure and body-part proportions. In our experience, gesture-generation researchers often experience significant friction when trying to transfer this skeletal data onto a realistic 3D character using animation software. For example, dataset releases do not include compatible skinned character models. This often means that an additional motion retargeting process is necessary for viewing the motion, which can introduce subtle artefacts in the resulting motion and requires expertise that adds barriers to enter the field.

To enable researchers to avoid the above issues, we are going to release carefully validated visualisation assets and pipelines as part of the leaderboard. We are currently developing an extended version of BVHView\cite{BVHView} for viewing gesturing animations on a 3D character during model development, released at \href{https://github.com/TeoNikolov/BVHView/}{github.com/TeoNikolov/BVHView/}. Our open-source tool allows users to visualise BVH files on the provided 3D avatar with synchronised speech audio in a simple drag-and-drop manner.
Additionally, we will develop an automated rendering pipeline for creating user-study video stimuli by either improving the Blender visualiser that was used during the GENEA challenges, or create a new tool using Unreal Engine or Unity. %

\subsection{Towards better objective metrics}\label{ssec:towards_better_metrics}
As discussed in \cref{sec:surveyed_object_evaluation}, automated performance metrics suffer from a lack of validity, and the core of the proposed leaderboard is to standardise and carry out subjective, human evaluation. That said, objective metrics can nonetheless offer guidance during system development and provide cost-effective ways to quantify concepts such as motion diversity. We will thus include objective performance metrics on the leaderboard and supply code for replicating the calculations behind them.

For all systems participating in the evaluation, we will publish a collection of their synthetic outputs, alongside all human responses gathered from the user studies. We consider this one of the most important potential contributions of the leaderboard, for several reasons. First, by having access to large amounts of synthetic outputs, independent researchers will be able to compare their own solutions to models on the leaderboard, even if there is no code available; compare with how \cite{ferstl2021expressgesture, yazdian2022gesture2vec} re-used stimuli published by the GENEA Challenge 2020. Second, the wide availability of synthesised gestures will enable important research on the effects of different evaluation methodologies and user-study interfaces; cf.\ the work in \cite{wolfert2021rate}. Additionally, raw human ratings may be used for further in-depth analysis on, for example, rating statistics, tendencies, or demographic effects; cf.\ \cite{kucherenko2024evaluating}. Finally, motion data paired with human ratings has immense potential to unlock the development of learnt objective evaluation metrics, as pioneered by \cite{he2022automatic} on GENEA-Challenge data.

\subsection{Community involvement}
\label{sec:community_involvement}
In contrast to the GENEA challenges, which were developed by a fixed group of organisers, community involvement has to be at the core of the GENEA Leaderboard project in order for it to succeed. This includes involvement in shaping the evaluation methodology, contributing resources like data and tooling for running the evaluations, as well as providing a wide array of systems to compare.

Developing a successful evaluation standard for gesture generation requires a broad understanding of the needs and goals of different research communities that work on or with gesture synthesis.
The plan described in this paper is not only based on discussions with senior researchers in gesture and its evaluation, but has been revised based on interaction with experts from the computer vision and machine-learning communities.
We are planning to establish an open community space (e.g., through GitHub) for %
discussions and providing additional opportunities to contribute. Such community contributions could address any part of gesture generation research: additional data, annotations, objective metrics, or even plugins for deploying gesture models in various environments, to name a few examples.

If the leaderboard successfully becomes the standard platform for evaluating gesture generation models, community contributions and participating systems will also achieve increased scientific impact. This is how we believe the leaderboard can align research incentives with better evaluation practices.

\subsection{Standardised resources}
In summary, beyond maintaining an up-to-date ranking of recent models, the GENEA Leaderboard will release %
resources supporting the full pipeline of developing and assessing gesture-generation models. We commit to releasing the following resources as part of the leaderboard project:
\begin{enumerate}
    \item A version of a selected \textbf{benchmark dataset} that is carefully adapted to the evaluation needs of the community and leaderboard. 
    \item One or more end-to-end \textbf{scene-rendering pipelines} that can be used both for visually inspecting synthesised gestures on a skinned 3D humanoid mesh during model development, and for creating canonical user-study stimulus videos for subjective evaluations.
    \item All \textbf{synthetic motion clips} submitted to the evaluation.
    \item Scripts to compute the most informative \textbf{objective metrics} in a standardised manner.
    \item A database of all \textbf{collected human responses} from the user studies.
    \item Collected \textbf{scripts for automating human evaluation}. 
\end{enumerate}
We believe these standardised and open-source resources will play a crucial role in making evaluations reproducible and, ultimately, contribute to comparing technologies and measuring progress. %
\color{black}
\section{Towards functional evaluations}
\label{sec:new_directions}
So far in this paper, we have exclusively focused on the evaluations used in the field today, and how these can be standardised and made easier to perform through our leaderboard, with the aim to improve the coverage of these evaluations to encompass all relevant gesture-generation research. 

However, %
once the use of the GENEA Leaderboard becomes widespread in the community, %
we intend to use it as a platform for incentivising research specifically to overcome important and long-standing shortcomings of gesture generation. This section describes how.

\subsection{End goals of gesture generation}
Gesture-generation research is widely motivated by the importance of gesturing for both humans and artificial agents. Gestures enrich human communication in many ways: they convey meaning (redundant or non-redundant to speech), underline and punctuate speech, serve pragmatic functions, help regulate conversation, express affective or attitudinal stance towards what is being said, adhere to cultural norms, reflect cognitive processes, and more \cite{wagner2014gesture}. All of these may be possible to infer from a gesture and hence impact how it, and the gesturing agent, are being perceived and evaluated. 

We argue that measuring the extent to which synthetic gestures enhance communication should be the long-term goal of gesture generation, but this will require fundamental changes to how evaluation is conducted in the field. %
This is supported by an experiment by \citet{saund2021importance}, in which participants were asked to rate gesture clips from the Trinity Speech-Gesture Dataset II \cite{ferstl2021expressgesture} without audio, based on two criteria: how well they match the meaning of the provided text transcript of the speech, and how energetic the speaker looks in the video. Importantly, they found no statistically significant differences between evaluator preferences for real gestures versus randomly selected, mismatched gestures from the dataset. Furthermore, they observed a high correlation between the semantic appropriateness rating of the gesture, and the perceived energy level of the speaker. These results corroborate the findings of the first GENEA Challenge \cite{kucherenko2021large}, where the baseline of presenting mismatched clips (with high human-likeness) achieved high appropriateness scores, even in the presence of the speech audio track.

The communicative functions gestures serve are closely defined by the context and the goals of the interaction. Therefore, it is perfectly normal that the spontaneous, unrestricted monologues of the Trinity dataset do not prominently feature gestures that closely connect with the meaning of the speech, i.e., not to an extent that moves the needle in a standard evaluation paradigm. Unfortunately, other mocap datasets used for gesture generation are similarly limited, e.g., the Talking With Hands dataset \cite{lee2019talking} has a mixture of free conversations and video retellings, while the BEAT dataset \cite{liu2022beat} contains neutral conversations and emotional \emph{readings} of written scripts about everyday topics.
\begin{tcolorbox}[enhanced,attach boxed title to top center={yshift=-3mm,yshifttext=-1mm},
  colback=blue!5!white,colframe=blue!75!black,colbacktitle=white,coltitle=black,fonttitle=\bfseries,
  boxed title style={size=small,colframe=blue!75!black}]
To resolve this, we believe it is essential for gesture generation to depart from currently available open-ended speech-gesture corpora, instead focusing on recording new, targeted, domain-specific data.
\end{tcolorbox}
\noindent Such data can in turn support more deliberate and targeted evaluations. In particular, recorded conversations must be coupled with clearly defined conversational, social, or functional goals, which can then be directly used in the evaluation of synthetic gestures. 

In practice, gestures must be grounded in the \emph{context of the interaction} in order to facilitate communication. Therefore, besides the speech, additional grounding information needs to be included the in the dataset, which can serve as additional input modalities for the generative model \cite{nyatsanga_comprehensive_2023}. The exact nature of the expected grounding highly depends on the conversational setting, but we believe that modelling advances for increased grounding will generalise between different settings, just as advances in deep generative modelling have been found to generalise well between different modalities.
Therefore, we believe that highly contextualised datasets will be better than generic ``one-size-fits-none'' datasets for driving progress towards general improvements in gesture generation.

Current evaluations, that lack grounding, will not show whether gestures enhance communication.
In fact, we can see that top systems in recent GENEA challenges \cite{yoon2022genea,kucherenko2023genea,kucherenko2024evaluating} are rated close to the human-motion capture data in terms of human-likeness, and several papers show results compatible with human parity or better \cite{rebol2021passing,zhou2022gesturemaster}.
The corresponding GENEA speech-appropriateness evaluations do find a large gap between natural and synthetic gestures, but they otherwise generally fail to differentiate between the submitted systems.
It has even been argued by \citet{nyatsanga_comprehensive_2023} that current evaluation setups bias data-driven models towards producing ``\emph{marginally natural gestures that appear more like well-timed hand waving, are not communicative and have little meaning}''.

\subsection{Contexts and methods for better evaluations}
We now propose a selection of possible interaction goals and contextualised datasets for gesture generation, and discuss novel ways for performing in-context evaluations on synthetic gestures. We note that the proposed, contextualised datasets might be challenging to collect, and might thus be limited in size. Embracing them for training and evaluation may therefore also create a much-needed push towards developing approaches that learn communicative gesturing behaviour across multiple domains, e.g., using pre-training, semi-supervised learning, or transfer-learning techniques. 

\paragraph{Conceptual understanding and recall:}
One way to measure whether gestures successfully enhance communication is to assess whether they improve the understanding of complex reasoning, or recalling what was said (similar to \cite{beattie2001experimental} or \cite{freigang2017pragmatic}, or the story-recall test in \cite{lin2021gestures}). 
This evaluation would be well-supported by a dataset containing conversations in an educational setting, or interactions that involve multistep decision-making processes. An important quality of  synthetic gesturing that can be evaluated here is if information that is contributed only by gesture (and \textit{not} by speech) is successfully picked up, integrated with speech meaning, and recalled later.

\paragraph{Intelligibility:}
Gestures also play a crucial role in improving speech comprehensibility in the presence of real-world background noise \cite{holle_integration_2009, wilms_effects_2022}. %
By adding artificial or even natural occurring noise to a dataset, proposed co-speech gesture-generation algorithms could be evaluated on their level of gesture intelligibility (e.g., disambiguating between different possible keywords or concepts from the message) under a realistic representation of actual communication challenges.%
Since consistently generating semantically 
expressive gestures is likely too challenging for open-domain conversation, we suggest focusing the evaluation and data collection effort on a limited set of concepts. 

\paragraph{Spatial referent disambiguation:}
In shared physical (or possibly virtual) environments, spatial gestures (including but not limited to pointing) can significantly enhance communication. These represented about 70\% of the gestures in a study of a conversational task grounded on a floor plan~\cite{smith2018communication}.  Spatial referencing gestures can be grounded in the spatial layout of the scene, and the location of objects in the environment. The communicative efficiency of gestures may be measured by the listener's ability to identify the referent object among plausible alternatives, as done by~\citet{deichler2023learning}. Another type of evaluation could feature agents guiding human participants via spoken instructions to solve tasks where object use is vital; robot assistants that guide cooking may be one concrete example of this. In this case, successful task completion rate could form the basis of a functional evaluation.

\paragraph{Creative control:} In creative media, expressing a certain personality or emotional state for the character via gestures can be more important than conveying meaning. To capture a broad range of emotions, gesture generation could be grounded by the intended valence and arousal levels \cite{russell1980circumplex}, which would have to be annotated during data collection. An alternative, more flexible solution is to express the desired gesturing style via an example motion recording, which would be provided as additional input to an exemplar-based generative model \cite{ghorbani2023zeroeggs}. In this case, with the appropriate dataset, one potential evaluation could assess the recognisability of different attributes---age, gender, and so on---from the gestures.

\paragraph{Multi-party conversations:} Moreover, capturing the dynamics of multi-party conversations in datasets provides insights into the intricacies of gestural interactions between multiple speakers. Although some datasets in this direction already exist, new, more comprehensive data sets could help in deciphering important aspects both for co-speech gesture generation, but also advance our knowledge in turn-taking behaviour and the overlap of speech and gestural synchrony between participants. Evaluation metrics can include measures of gestural coherence in a given context and alignment across speakers within the conversation.

\paragraph{}

\section{Conclusion}
In this paper we take the position that the evaluation of speech-driven gesture generation needs to improve in order to support comparing proposed methods to each other and, especially, to drive and measure progress towards the goals that motivate research in the field.
Our argumentation is supported by a survey of recent evaluations of gesture-generation systems in computer vision and graphics, and by a synthesis of prior literature and findings.

To overcome the limitations of current evaluation practices, we announced the GENEA Leaderboard project, a living benchmark for conversational motion synthesis.
We detailed how this initiative can benefit the field, first by standardising and streamlining evaluations around best practises -- for improved comparison coverage and reproducibility -- and in the longer-term as an evolving platform for enabling evaluations on more meaningful data and tasks that promote future research towards the goals of the field. 

Community involvement is central to the success of the proposed leaderboard, and we encourage input and contributions from readers, in addition to our active solicitation of input from research-community members.

\section{Acknowledgements}
We thank Tiago Ribeiro and Saeed Ghorbani for providing valuable feedback on the project. This work was partially supported by the Wallenberg AI, Autonomous Systems and Software Program (WASP) funded by the Knut and Alice Wallenberg Foundation and by the Industrial Strategic Technology Development Program (grant no.\ 20023495) funded by MOTIE, Republic of Korea.

{
    \small
    \bibliographystyle{ieeenat_fullname}
    \bibliography{main}
}

\end{document}

%% file: preamble.tex
\makeatletter
\robustify\@latex@warning@no@line
\makeatother
\usepackage{authblk}
\usepackage[dvipsnames]{xcolor}
\usepackage[most]{tcolorbox}
\usepackage{multirow}
\usepackage{makecell}
\usepackage{subcaption}
\usepackage{color,soul}
\usepackage{tikz}
\usetikzlibrary{decorations.pathreplacing,calc,arrows,automata}

\newcommand{\tikzmark}[2][-3pt]{\tikz[remember picture, overlay, baseline=-0.5ex]\node[#1](#2){};}

\usepackage[dvipsnames]{xcolor}

%% file: reviewed_works.tex
\newcounter{brace}
\setcounter{brace}{0}
\newcommand{\drawbrace}[3]{%
 \refstepcounter{brace}
 \tikz[remember picture, overlay]
 \fill ($(#2.center)+(#3,0)$) circle[radius=2pt];
 \tikz[remember picture, overlay]\draw
 ($(#1.center)+(#3,0)$)--
 ($(#2.center)+(#3,0)$);
 \tikz[remember picture, overlay]
 \fill[blue] ($(#1.center)+(#3,0)+(-1.5pt,-1.5pt)$) rectangle++(3pt, 3pt);
 }

\begin{table}[!t]
    \centering
    \begin{tabular}{@{}rlrl@{}}
    \toprule
    Year & Conference & Model name & User studies\\
    \midrule

\multirow[t]{7}{*}{2022} & \multirow[t]{3}{*}{CVPR} & DiffGAN \cite{ahuja2022low} & \tikzmark[xshift=0em]{DiffGAN} \\
 &  & HA2G \cite{liu2022learning} & \tikzmark[xshift=0em]{HA2G} \\
 &  & SEEG \cite{liang2022seeg} & \tikzmark[xshift=0em]{SEEG} \\
 & SIGGRAPH & cGAN-KNN \cite{habibie2022motion} & \tikzmark[xshift=0em]{cGAN-KNN} \\
 & \multirow[t]{2}{*}{ECCV} & CaMN \cite{liu2022beat} & \tikzmark[xshift=0em]{CaMN} \\
 &  & FlowGesture \cite{ye2022audio} & \tikzmark[xshift=0em]{FlowGesture} \\
 & \multirow[c]{2}{*}{\shortstack[r]{SIGGRAPH\\Asia}} & \multirow[c]{2}{*}{\shortstack[r]{Rh. Gest. \cite{ao2022rhythmic}}} & \tikzmark[xshift=0em]{RhGest} \\
&&&\\
\multirow[t]{9}{*}{2023} & \multirow[t]{4}{*}{CVPR} & DiffGesture \cite{zhu2023taming} & \tikzmark[xshift=0em]{DiffGesture} \\
 &  & QPGesture \cite{yang2023qpgesture} & \tikzmark[xshift=0em]{QPGesture} \\
 &  & RACER \cite{sun2023co} & \tikzmark[xshift=0em]{RACER} \\
 &  & TalkSHOW \cite{yi2023generating} & \tikzmark[xshift=0em]{TalkSHOW} \\
 & \multirow[t]{3}{*}{SIGGRAPH} & Bodyformer \cite{pang2023bodyformer} & \tikzmark[xshift=0em]{Bodyformer} \\
 &  & GestureDiffuCLIP \cite{ao2023gesturediffuclip} & \tikzmark[xshift=0em]{GestureDiffuCLIP} \\
 &  & LDA \cite{alexanderson2023listen} & \tikzmark[xshift=0em]{LDA} \\
 & \multirow[t]{2}{*}{ICCV} & C-DiffGAN \cite{ahuja2023continual} & \tikzmark[xshift=0em]{C-DiffGAN} \\
 &  & LivelySpeaker \cite{zhi2023livelyspeaker} & \tikzmark[xshift=0em]{LivelySpeaker} \\
\multirow[t]{8}{*}{2024} & \multirow[t]{7}{*}{CVPR} & AMUSE \cite{chhatre2024amuse} & \tikzmark[xshift=0em]{AMUSE} \\
 &  & Audio2Photoreal \cite{ng2024audio2photoreal} & \tikzmark[xshift=0em]{Audio2Photoreal} \\
 &  & ConvoFusion \cite{mughal2024convofusion} & \tikzmark[xshift=0em]{ConvoFusion} \\
 &  & DiffSHEG \cite{chen2024diffsheg} & \tikzmark[xshift=0em]{DiffSHEG} \\
 &  & EMAGE \cite{liu2024emage} & \tikzmark[xshift=0em]{EMAGE} \\
 &  & EmoTransition \cite{qi2024emotransition} & \tikzmark[xshift=0em]{EmoTransition} \\
 &  & ProbTalk \cite{qi2024emotransition} & \tikzmark[xshift=0em]{ProbTalk} \\
 & SIGGRAPH & Sem. Gest. \cite{zhang2024SemanticGesture} & \tikzmark[xshift=0em]{SemGest} \\

\bottomrule
\end{tabular}
\caption{List of papers considered for the analysis in \cref{sec:current_state}. In the last column, blue squares indicate that the authors used human evaluation to compare to other models in the table, which are represented as connected black dots.}
\label{tab:recent_models}
\vspace{-1\baselineskip}
\end{table}

\drawbrace{SemGest}{GestureDiffuCLIP}{0.0em}
\drawbrace{SemGest}{CaMN}{0.0em}
\drawbrace{EmoTransition}{CaMN}{0.5em}
\drawbrace{EmoTransition}{HA2G}{0.5em}
\drawbrace{EmoTransition}{DiffGesture}{0.5em}
\drawbrace{EMAGE}{TalkSHOW}{1.0em}
\drawbrace{DiffSHEG}{CaMN}{1.5em}
\drawbrace{DiffSHEG}{DiffGesture}{1.5em}
\drawbrace{DiffSHEG}{LDA}{1.5em}
\drawbrace{DiffSHEG}{TalkSHOW}{1.5em}
\drawbrace{ConvoFusion}{CaMN}{2.0em}
\drawbrace{Audio2Photoreal}{LDA}{2.5em}
\drawbrace{AMUSE}{TalkSHOW}{3.0em}
\drawbrace{LivelySpeaker}{HA2G}{3.5em}
\drawbrace{C-DiffGAN}{DiffGAN}{4.0em}
\drawbrace{GestureDiffuCLIP}{CaMN}{4.5em}
\drawbrace{QPGesture}{cGAN-KNN}{5.0em}
\drawbrace{QPGesture}{CaMN}{5.0em}
\drawbrace{DiffGesture}{HA2G}{5.5em}

%% file: roadblocks_sections.tex
\renewcommand{\arraystretch}{1.2}
\begin{table}[]
    \centering
    \begin{tabular}{ll}

\toprule
Roadblocks for better evaluations & Addressed by  \\
\midrule
No definitive benchmark dataset & \cref{sec:benchmark_dataset} \\
\multirow[c]{2}{*}{\shortstack[l]{Lack of standardised user-study designs\\and study-creation tools}} & \multirow[t]{2}{*}{\cref{sec:user_study_design}} \\
&\\
\multirow[c]{2}{*}{\shortstack[l]{Undocumented test-set selection process\\in user studies}} & \multirow[t]{2}{*}{\cref{ssec:segment_selection}} \\
&\\
Inconsistent visualisations in evaluations & \cref{ssec:visualisation_tooling} \\
\multirow[c]{2}{*}{\shortstack[l]{Reimplemented or newly trained baselines\\ are often unreliable}} & \multirow[t]{2}{*}{\cref{ssec:initial_seeding}} \\
&\\
Lack of reliable objective metrics & \cref{ssec:towards_better_metrics} \\
\bottomrule   
    \end{tabular}
    \caption{The GENEA Leaderboard initiative is a holistic solution for the evaluation roadblocks identified in \cref{sec:current_state}.}
    \label{tab:roadblocks_sections}
\end{table}
\renewcommand{\arraystretch}{1}